# Synthetic Voice Spoofing Detection Based On Online Hard Example Mining

Chenlei Hu, Ruohua Zhou

The automatic speaker verification spoofing (ASVspoof) challenge series is crucial for enhancing the spoofing consideration and the countermeasures growth. Although the recent ASVspoof 2019 validation results indicate the significant capability to identify most attacks, the model's recognition effect is still poor for some attacks. This paper presents the Online Hard Example Mining (OHEM) algorithm for detecting unknown voice spoofing attacks. The OHEM is utilized to overcome the imbalance between simple and hard samples in the dataset. The presented system provides an equal error rate (EER) of 0.77% on the ASVspoof 2019 Challenge logical access scenario's evaluation set.

*Introduction:* Automated speaker authentication systems, which identify speakers by voice, are now widely utilized in engineering applications [1]. However, with recent advances in artificial intelligence and hardware technology, ASV systems are threatened by various voice spoofing attacks. Illegals attack ASV systems and deceive humans by creating spoof voices such as mimic, replay, and synthetic voice attacks (involving text-to-speech (TTS) and voice conversion (VC)) to impersonate real users [2]. Reliable anti-spoof systems are required to alleviate or eliminate the risk of fraud to ASV systems and human users.

Most spoofing countermeasure systems comprise the front and back end [3]. The traditional front ends are based on digital signal processing algorithms, such as linear frequency cepstral coefficients (LFCC)[4] and Constant-Q transform (CQT)[5], to extract acoustic features. Besides, some studies indicate that more discriminative acoustic features can be extracted for anti-spoofing tasks by adding front-end feature extraction to the training model. Similarly, DNN determines the filter's center frequency in the filter bank [6]. As for the back ends, many studies have employed convolutional neural networks and loss functions [7] for face verification and image classification tasks. For example, Galina Lavrentyeva et al. [8] utilized a novel convolutional network, LCNN, for replay and synthetic speech identification. Li et al. [9] applied a new variant of the residual network to replay and synthetic spoof detection.

The previous systems cannot generalize the unseen spoofing attacks in the evaluation step [10]. This paper employs OHEM to develop an anti-spoofing system to distinguish unknown synthetic voice spoofing attacks. OHEM with screened sampling is a commonly used sampling approach, which selects the training examples to promote the anti-spoofing system's efficiency. This work focuses on combining the OHEM algorithm with the anti-spoofing model.

This paper is organized as the following. The application of the OHEM algorithm in synthetic speech spoofing detection is described in Section 2. The details of the experimental design and results are given in Section 3. Section 4 concludes the paper.

*Methods:* For network training, the objective function directly defines the latent mappings that the network should fit. The quality of the samples counted to minimize the objective function can determine the realization degree of the network's goal. This work indicates that the number of negative samples is larger than that of positive ones in a limited training set sample, which can bias the loss function. Besides, most negative samples are easily classified. The ratio of simple samples is much greater than that of hard samples. Moreover, the loss of simple samples is much lower than the hard ones (around zero).

Therefore, an online hard negation example mining (OHEM) anti-spoofing model is proposed, which employs the OHEM strategy to tune the objective function and selectively search informative hard negative samples to improve the training efficiency significantly. Motivated by [11], the proposed OHEM strategy is based on excluding non-informative samples from the loss during the training to alleviate the imbalance between simple and hard samples. In this regard, the proposed OHEM comprises three main steps in each training iteration. First, the number of hard negative examples should be selected. This number is adaptively determined as N/4, where N represents the number of samples in the mini-batch. Then, the N training samples are sorted according to their prediction scores. Finally, our loss function computes only the top N/4 sample losses and discards the rest. We ignore simple negative samples and only focus on the hard samples.

*Research process*：This section describes the ASVspoof2019 LA datasets, front-end features, and back-end network models used in the experiments. Besides, four sets of experimental results and the fusion system results are given.

A. Datasets

All tests were performed using the ASVspoof 2019 Logical Access (LA) database. Table 1 presents a detailed representation of the two mentioned subsets. Training and Developing sets share similar six attacks. The six attacks mainly contain two VC and four TTS algorithms. In the evaluation set, there are 11 unknown attacks (A07-A15, A17, A18) including combinations of different TTS and VC attacks. The evaluation set also includes two attacks (A16, A19) which use the same algorithms as two of the attacks (A04,A06) in the training set but were trained with different data. [10].

**Table 1:** ASVspoof 2019 LA Datasets

| Dataset | Bona fide | Spoofed | |
|---|---|---|---|
| | utterance | utterance | attacks |
| Training | 2580 | 22800 | A01-A06 |
| Developing | 2548 | 22296 | A01-A06 |
| Evaluation | 7355 | 63882 | A07-A19 |

B. Feature selection

For manual feature selection, the code prepared by the official ASVspoof 2019 competition was employed for extracting 60-dimensional LFCCs features from the raw voice data, containing a 20ms window size and a 10ms sliding window. For the raw audio, the data were fixed to a voice length of about 4 seconds to uniformize the size of all utterances.

C. Model selection

The experiments involve two models. One is based on the residual network variant [12], containing Resnet-18, Resnet-50, and SE-res2net [9]. The other is a new variant proposed based on the rawnet2 [8] architecture. The residual network has been utilized with excellent results and applications in anti-spoofing. Table 2 shows the detail, where the first input is the raw audio. The SincNet layer is passed first, followed by the residual structure. Besides, the original residual block structure for the residual block is replaced with the 1D Res2net block. Finally, the GRU operation is performed before the full connection.

**Table 2:** The architecture of Raw-res2net, BN refers to batch normalization. Cons involves the block convolution of the Res2net architecture and the BN&LeakyReLu operation, and SELayer stands for squeezing excitation block.

| Layer | Input:**64600** samples | Output shape |
|---|---|---|
| **Fixed** Sinc filters | Conv (**1024**,1,20)<br>Maxpooling(3)<br>BN & LeakyRelu | (**21192**,20) |
| **SE-Res2net block (*2)** | BN & LeakyReLU<br>Conv (1,1,20)<br>BN & LeakyReLU<br>Conv (3,1,20)<br>Conv (1,1,20)<br>BN & LeakyReLU<br>Maxpooling(3)<br>SELayer | (**2354**,20) |
| **SE-Res2net block (*4)** | BN & LeakyReLU<br>Conv (1,1,**128**)<br>BN & LeakyReLU<br>Convs(3,1,**128**)<br>Conv (1,1,**128**)<br>BN & LeakyReLU<br>Maxpooling(3)<br>SELayer | (**29**,128) |
| GRU | GRU (1024) | (1024) |
| FC | 1024 | (1024) |
| **Output** | **1024** | **2** |

D. Results and Analysis

1) **OHEM effectiveness evaluation**: Four groups of experiments have been designed under the same experimental conditions. Table 3 presents the experimental results. The introduction of OHEM has led to an

EER reduction in all models. In particular, the Resnet18-OHEM model achieved a 42% reduction in EER, and the t-DCF index decreased by nearly 50%. The best result for the single model is Res2net-OHEM, achieving an EER of 2.13%. Fig. 1 compares the performance of Resnet18 and Resnet18-OHEM models for different logical attacks. As shown in Fig. 1, the Resnet18-OHEM system outperforms Resnet18 under most spoofing attacks (A07-A15, A17). Moreover, after the introduction of OHEM, the model has improved to different degrees for almost every attack identification.

2) **Fusion results and comparison with other systems**: Table 4 shows our scores for fusing different models and compares them with other models. After performing a two-by-two fusion of the models, it was discovered that the S3 model combined with other models could provide t-DCF values of 0.0233 and 0.0262, respectively. Analysis revealed that the S3 model outperforms at identifying A17, and this benefit may be fully realized by merging it with other models. Besides, EER=0.77 and t-DCF=0.0228 were derived by combining the results of the S1, S2, and S3 models.

**Table 3:** The comparison between the four models after the introduction of OHEM.

| Models | Resnet18 | | Res2net | | Raw-res2net | | Resnet50 | |
|---|---|---|---|---|---|---|---|---|
| | EER | t-DCF | EER | t-DCF | EER | t-DCF | EER | t-DCF |
| **OHEM** | **2.32** | **0.052** | **2.04** | **0.054** | 3.03 | 0.078 | 3.32 | 0.083 |
| Without OHEM | 3.99 | 0.105 | 2.73 | 0.075 | 3.88 | 0.107 | 4.58 | 0.109 |

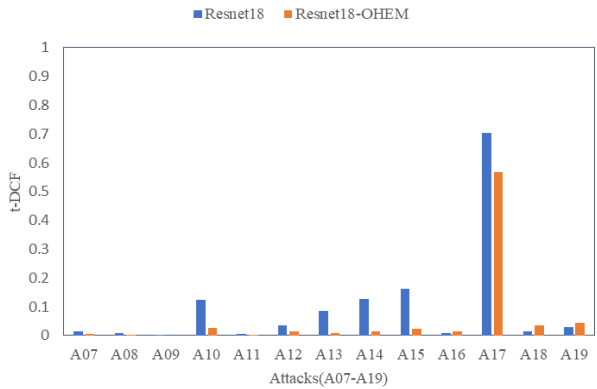

**Fig. 1** *Models in the evaluation set regarding (A07-A19) attack results.*

**Table 4:** Performance comparison between ASVspoof 2019 LA evaluation set and existing submission system. S1 represents the model Resnet18-OHEM, S2 represents the model Res2net-OHEM, and S3 represents the end-to-end model Raw-Res2net-OHEM. S1 + S2 + S3 and (S1 + S2, S2 + S3, S1 + S3) represent the score fusion of three models and the fusion of two models, respectively.

| System | t-DCF | EER |
|---|---|---|
| **S1+S2+S3** | **0.0228** | **0.77** |
| **S1+S3** | **0.0233** | **0.85** |
| **S2+S3** | **0.0262** | **1.02** |
| T45[8] | 0.0510 | 1.86 |
| **S1+S2** | **0.0455** | **1.95** |
| **S2** | **0.0547** | **2.04** |
| OC-softmax [10] | 0.0590 | 2.19 |
| **S1** | **0.0522** | **2.32** |
| SE-Res2net [9] | 0.0743 | 2.50 |
| **S3** | **0.0785** | **3.03** |
| Rawnet2 [8] | 0.1294 | 4.66 |

*Conclusion:* The current paper applies OHEM to speech anti-spoofing systems and experimentally employs four different sets of models. The experimental results indicate that the presented system can identify some unseen attacks well. Our best system achieved an EER of 0.77 % by a score-level fusion. For future work, we hope to employ the methods used in our experiments in more anti-spoofing models.

**References**
1. Delac K, Grgic M. A survey of biometric recognition methods[C]//Proceedings. Elmar-2004. 46th International Symposium on Electronics in Marine. IEEE, 2004: 184-193.
2. Wu Z, Evans N, Kinnunen T, et al. Spoofing and countermeasures for speaker verification: A survey[J]. speech communication, 2015, 66: 130-153.
3. Wang X, Yamagishi J. Investigating self-supervised front ends for speech spoofing countermeasures[J]. arXiv preprint arXiv:2111.07725, 2021.
4. Davis S, Mermelstein P. Comparison of parametric representations for monosyllabic word recognition in continuously spoken sentences[J]. IEEE transactions on acoustics, speech, and signal processing, 1980, 28(4): 357-366.
5. Todisco M, Delgado H, Evans N. Constant Q cepstral coefficients: A spoofing countermeasure for automatic speaker verification[J]. Computer Speech & Language, 2017, 45: 516-535.
6. Tak H, Patino J, Todisco M, et al. End-to-end anti-spoofing with rawnet2[C]//ICASSP 2021-2021 IEEE International Conference on Acoustics, Speech and Signal Processing (ICASSP). IEEE, 2021: 6369-6373.
7. Wang X, Yamagishi J. A comparative study on recent neural spoofing countermeasures for synthetic speech detection[J]. arXiv preprint arXiv:2103.11326, 2021.
8. Lavrentyeva G, Novoselov S, Tseren A, et al. STC antispoofing systems for the ASVspoof2019 challenge[J]. arXiv preprint arXiv:1904.05576, 2019.
9. Li X, Li N, Weng C, et al. Replay and synthetic speech detection with res2net architecture[C]//ICASSP 2021-2021 IEEE international conference on acoustics, speech and signal processing (ICASSP). IEEE, 2021: 6354-6358.
10. Zhang Y, Jiang F, Duan Z. One-class learning towards synthetic voice spoofing detection[J]. IEEE Signal Processing Letters, 2021, 28: 937-941.
11. Shrivastava A, Gupta A, Girshick R. Training region-based object detectors with online hard example mining[C]//Proceedings of the IEEE conference on computer vision and pattern recognition. 2016: 761-769.
12. He K, Zhang X, Ren S, et al. Deep residual learning for image recognition[C]//Proceedings of the IEEE conference on computer vision and pattern recognition. 2016: 770-778.